\documentclass[submission,copyright,creativecommons]{eptcs}
\providecommand{\event}{SOS 2007} 
\usepackage{breakurl}             %
\usepackage{underscore}           
\usepackage{graphicx}

\title{The Forms of Categorical Proposition}
\author{Fabien Schang
	\institute{UFG\\ Goiânia, Brazil}
	\email{schangfabien@gmail.com}
	\and
	George Englebretsen 
	\institute{Bishop's University\\ Sherbrooke, Canada}
	\email{genglebr@ubishops.ca}
	\and J.-Mart\'in Castro-Manzano
	\institute{UPAEP University\\
		Puebla, Mexico}
	\email{josemartin.castro@upaep.mx}
}
\def\titlerunning{The Forms of Categorical Proposition}
\def\authorrunning{F. Schang, G. Englebretsen \& J.\,M. Castro Manzano}
\begin{document}
	\maketitle
	
	\begin{abstract}
		An exhaustive survey of categorical propositions is proposed in the present paper, both with respect to their nature and the logical problems raised by them. Through a comparative analysis of Term Logic and First-Order Logic, it is shown that the famous problem of existential import may be solved in two ways: with a model-adaptive strategy, in which the square of opposition is validated by restricting the models; with a language-adaptive strategy, in which the logical form of categorical propositions is extended in order to validate the square in every model. The latter strategy is advocated in the name of logic, which means truth in every model. \noindent Finally, the present paper needs some automatic process in order to determine the nature of logical relations between any pair of the available 256 categorical propositions. This requires the implementation of a programming machine in the style of Prolog.
	\end{abstract}
	
	\section{What is a categorical proposition?}
	It is taken for granted that the roots of modern logic stem from Aristotle's syllogistics, as a general inferential activity. However, the content of such inferences was initially restricted to a specific kind of formulas: \textit{categorical propositions}. A general definition of the latter is the following: by a categorical proposition, it is meant
	any proposition in which something, say $A$, is affirmed or denied of something,
	say $B$ ($A$ and $B$ may be two names of the same thing).\footnote{Although, literally speaking, a categorical proposition also includes an expression of \textit{quantity} (in order to say how many elements of the $A$ are $B$), the above definition may be viewed as a categorical \textit{proto-proposition} (see \cite{Eng80}), i.e. the core expression from which any categorical proposition and its extensions may be formed. We are grateful to one of the anonymous reviewers for that relevant note about what categoricity traditionally means.} Every thought may be
	expressed in that way, provided that the grammatical form of the proposition includes a first term, the subject, and a second term, the predicate.
	
	\noindent The aim of the following is not to deal with syllogistics as it stands; rather, we want to deal with the various ways in which categorical propositions may be analyzed, especially in the light of their logical form. A number
	of questions may arise from this perspective: ‘Every $A$ is $B$’ is a typical categorical proposition, whereas ‘Everything is $A$ and $B$’ doesn’t seem so.
	Isn’t it? Does ‘Every $A$ is $B$’ entail that ‘Some A is B’? What is the difference between ‘$A$ is not-$B$’ with respect to ‘$A$ is not $B$’? Here are some issues that
	are tackled in the following work, in order to give a comprehensive survey of what is meant by the so-called ‘categorical propositions’.
	
	\noindent A basic proposition is of the form ‘$A$ is $B$’ and constitutes the core expression of the so-called traditional Term logic (henceforth: TL)(see \cite{Eng16b}). According to TL, every proposition is a pair of terms bound together by a (logical) copula to
	form a single unified expression. Modern logic results from a choice: for a number of reasons, Frege defeated Aristotle and the contemporary First-Order Logic (henceforth: FOL) is the mainstream formal language that is mainly used in mathematics and the other formal sciences. Assuming that the value of a language lies in its expressive power, FOL is expected to be able to express as much of our thoughts as Term Logic. This is not so easy in any case, however. Let us
	consider \newline
	
	\noindent (1) Ryan is even. \newline
	
	\noindent (1) is expected to be false, for ‘even’ is to be predicated of whatever is a number only. So the propositional negation of (1), \newline
	
	\noindent (1)* Ryan is not even, \newline
	
	\noindent is true by \textit{denying} a predicate that does not belong to the properties of whatever is not a number. But, as we will see in the next section, there are other negative expressions that are not applied to entire propositions and occur inside the propositions, e.g. the affixal negations like ‘undefeated’ or ‘impossible’. Let us call \textit{term negations} those cases in which negation is applied only to one term of the proposition (e.g. ‘defeated’ and ‘possible’, in the previous two examples), to be symbolized by a hyphen in order to make the difference between propositional negation in ‘$A$ is not $B$’ and ‘$A$ is not-$B$.\footnote{This is confirmed by Aristotle in his \textit{Prior Analytics}, I,46: ``It is clear that ‘is not-white’ and ‘is not white’ signify different things and that one is an affirmation, the other a denial.'' (52a24-26). (Also quoted in Read 2015: 9).} Traditional logic made a distinction between propositional negation and term negation, accordingly: (1)* means something like ‘It is not the case that Ryan is even’, whereas \newline
	
	\noindent (2) 9 is not-even \newline
	
	\noindent is true by \textit{affirming} that the number 9 is not-even, that is, odd.
	
	\noindent FOL is not able to express term negation, because the traditional copula finds no expression in it and this reduces negation to a unique propositional
	operator. Whilst (1)* is true, the other proposition \newline
	
	\noindent (1)$'$ Ryan is not-even \newline
	
	\noindent is false since being not-even is synonymous with being odd. TL makes this
	distinction in terms of distinctive \textit{oppositions} (see \cite{Eng84}, \cite{Eng16a}): ‘$A$ is $B$’ and ‘$A$ is not $B$’ are said to be contradictories because the truth of the one entails the falsehood
	of the other and conversely; at the same time, ‘$A$ is $B$’ and ‘$A$ is not-$B$’ are said to be \textit{contraries} because the truth of the one entails the falsehood of the other whilst both can be false (Ryan is neither even nor odd, indeed).
	
	\noindent Although FOL lacks expressive power in this respect, it may neglect this
	point by recalling that formal logic essentially deals with generality and is not interested with individual truths (see \cite{Eng86}). Aristotle’s logic confirms this point by giving priority to general propositions about classes of individuals, like the $A$s or the $B$s. As FOL is able to rephrase every such proposition, the previous
	semantic distinction between (1)* and (1)$'$ is ignored. Nonetheless, let us now consider \newline
	
	\noindent (3) Dragons are dangerous. \newline
	
	\noindent (3) is admittedly said true in most of its fictional statements. But the corresponding proposition is supposed to be false for whomever has a robust sense of reality, because dragons do not exist. So the contradictory of (3) is true, i.e. (3)* Dragons are not dangerous, by definition. Then what of its presumed contrary, i.e. \newline
	
	\noindent (3)$'$ Dragons are not-dangerous, \newline
	
	\noindent which affirms something and means that dragons are not-dangerous, that is, friendly? (3)$'$ is as false as (3), if dragons do not exist. This is a way to say that the Aristotelian square of opposition includes some ambiguity behind its usual headings. The proposition ‘Every $A$ is $B$’ is called \textit{affirmative universal}, because being $B$ is affirmed of every $A$, whereas the proposition ‘Every $A$ is not $B$’ is called \textit{negative universal} because being $B$ is denied of every $A$. Then ‘Every $A$ is not-$B$’ should be considered as a special case of affirmative universal, in which the negative predicate not-$B$ is affirmed of every $A$. The point is that there is no way to express the difference between ‘Every $A$ is
	not $B$’ and ‘Every $A$ is not-$B$’ in FOL, or it seems so.
	
	The following purports to present both a syntax and a semantics for all
	these kinds categorical propositions without exception. In Section 2, a variety of logical forms is listed from proto-categorical to apodictic propositions. In Section 3, their truth-conditions are expressed in terms of ordered model sets.
	
	\section{Syntax}
	
	A clear-cut difference is to be made between two kinds of speech act, namely: affirming a negative predicate term $A$, ‘is not-$A$’, and denying a positive predicate $A$, ‘is not $A$’. The latter corresponds to what Aristotle called the \textit{composition} of a proposition in its ultimate stage. That is: asserting $A$ of something means that the resulting proposition is taken to be \textit{true}, whereas denying $A$ means that the resulting proposition is taken to be \textit{false}. The truth-conditions of propositions will be studied in
	the next section. Here is a set of increasingly complex logical forms of propositions, $\mathcal{A}^{n}_{C}$, where $n$ stands for the number of \textit{literals}, i.e. components of a compound formula that can either affirmed (+) or denied (–). The logical complexity of propositions depends on the number of literals that are in it. \newline
	
	\noindent Type 1: ‘$A$ is (not) $B$’.
	
	\noindent $\mathcal{A}^{1}_{C}$ = $A$ $\pm$is $B$
	
	\noindent $2^{1} = 2$ kinds of proposition: (1) $A$ is $B$, (2) $A$ is not $B$. \newline
	
	\noindent Type 2. ‘all/some $A$ is (not) $B$’. 
	
	\noindent $\mathcal{A}^{2}_{C}$ = $\pm$some $A$ $\pm$is $B$.
	
	\noindent $2^{2} = 4$ kinds of proposition: (1) some $A$ is $B$, ..., (4) all $A$ is not $B$. \newline
	
	\noindent Type 3. ‘(not-)$A$ is (not) (not-)$B$’. 
	
	\noindent $\mathcal{A}^{3}_{C}$ = $\pm A$ $\pm$is $\pm B$
	
	\noindent $2^{3} = 8$ kinds of proposition: (1) $A$ is $B$, ..., (8) not-$A$ is not not-$B$.\newline
	
	\noindent Type 4. ‘(all/some) (not-)$A$ is (not) (not-)$B$’. 
	
	\noindent $\mathcal{A}^{4}_{C}$ = $\pm$some $\pm A$ $\pm$is $\pm B$
	
	\noindent $2^{4} = 16$ kinds of proposition: (1) some $A$ is $B$, ..., (16) all not-$A$ is not not-$B$.\newline
	
	\noindent Type 5. (all/some) (not-)$A$ is (not) (necessarily/possibly) (not-)$B$
	
	\noindent $\mathcal{A}^{5}_{C}$ = $\pm$some $\pm A$ $\pm$is $\pm$possibly $\pm B$
	
	\noindent $2^{5} = 32$ kinds of proposition: (1) some $A$ is possibly $B$, ..., (32) all not-$A$ is
	necessarily not-$B$. \newline
	
	\noindent The rest of the paper will be confined to dealing with assertive propositions, i.e. the set of categorical propositions including Aristotle’s four copulae.
	Given that this class of proposition includes $n = 4$ literals in the syntax of TL, we will symbolize henceforth any categorical proposition by $A^{4}_{C}$ accordingly.
	
	In light of FOL, we take for the time being the following propositional
	scheme to be the common logical form of any categorical propositions: \begin{center}
		$\mathcal{A}^{4}_{C} = \pm$$\exists x(\pm$$Ax \wedge \pm$$Bx)$
	\end{center}
	
	\noindent The 3 literals that can be either affirmed or denied, viz. existential quantifier and the two predicates, lead to a set of $2^{3} = 8$ distinct formulas, i.e. 4
	categorical propositions $X$ with only affirmed subject terms and 4 categorical propositions $X'$ with only denied subject terms.
	
	Given that the propositional negation of the latter is logically equivalent
	to a universal, the following table affords a set of two possible formalization for categorical propositions: either in FOL, or in TL.
	
	\begin{center}
		$\begin{array}
			{|c|c|c|} \hline & FOL & TL \\
			\hline
			\textbf{A} & \neg(\exists x)(Ax \wedge \neg Bx) & -(+A-(+B)) \\ \hline 
			\textbf{E} & \neg(\exists x)(Ax \wedge Bx) & -(+A+B)) \\ \hline
			\textbf{I} & (\exists x)(Ax \wedge Bx) & +(+A+B) \\ \hline
			\textbf{O} & (\exists x)(Ax \wedge \neg Bx) & +(+A+(-B)) \\ \hline
			\textbf{A}' & \neg(\exists x)(\neg Ax \wedge Bx) & -(+A+B) \\ \hline
			\textbf{E}' & \neg(\exists x)(\neg Ax \wedge \neg Bx) & -(-A+(-B)) \\ \hline
			\textbf{I}' & (\exists x)(\neg Ax \wedge \neg Bx) & +(-A+(-B)) \\ \hline
			\textbf{O}' & (\exists x)(\neg Ax \wedge Bx) & +(-A+B) \\ \hline
		\end{array}$  \end{center} 
	
	Although Aristotle’s logical works were usually devoted to scientific discourse and arguably led to some disregard for singular propositions, we have seen that the latter may be analyzed on their own for purely logical reasons. In the same vein, there is a semantic reason why the Aristotelian theory of categorical propositions has usually been restricted to the set \{\textbf{A},\textbf{E},\textbf{I},\textbf{O}\}, whereas the other formulas \{\textbf{A}$'$,\textbf{E}$'$,\textbf{I}$'$,\textbf{O}$'$\} were always left out of consideration.
	
	\noindent So let us consider this semantics, before turning to a general consideration
	of categorical propositions and their proper logical form.
	
	\section{Semantics}
	
	Our main concern will be with models. \textit{Logical truth} means that a proposition is supposed to be true in every model, whilst a \textit{logical consequence} means that models of one formula are also models of entailed formulas. Although these general features of logic are assumed to hold in every model, some convenient
	restrictions are generally made in order to warrant the completeness of logical systems. One of these is the non-emptiness of models: What is the truth of ‘Every $x$ is $x$’ in an empty model, i.e. the model in which there is nothing?
	Whilst the corresponding formula $(x)(x = x)$ refers to the law of self-identity, it is equally true to affirm ‘No $x$ is $x$’ in the empty model, so that the contrary formula $\neg(\exists x)(x = x)$ is also true in it and, by classical logic,
	$(x)(x \neq x)$. Incompleteness is on a par with inconsistency, accordingly: classical logic
	cannot cope with emptiness. The same difficulty arises in modal logic, insofar as necessary and impossible truths are logically equivalent as well in the empty model. Does it follow from it that any logical system should always prohibit emptiness and make some minimal existential assumption for its own
	sake? Let us consider what happens in the contrary case, with the special case of categorical propositions and traditional logic.
	
	\subsection{Existential import}
	
	A semantics for categorical propositions is expected to afford the truth-conditions relative to a set of models these propositions belong to. Let us call ‘model set’ all those models where a given proposition is true. A model is a set of propositions including, e.g., $\varphi$, that are true in $w{i}$ if, and only if (henceforth: iff), they belong to $w_{i}$: $v(w_{i}, \varphi) = 1$ iff $\varphi\in w_{i}$; and they are false in $w_{i}$, otherwise. We write $v(w_{i}, \varphi) = 1$ for ‘$\varphi$ is true in $w_{i}$’, and $v(w_{i}, \varphi) = 0$ for ‘$\varphi$ is false in $w_{i}$’. As categorical propositions $\mathcal{A}^{n}_{C}$ are used to be analyzed into a bivalent frame, their corresponding models are expected to be consistent and complete. That is: propositions are either true or false in any model: $v(w_{i}, \mathcal{A}^{n}_{C}) = 1$ iff $v(w_{i}
	,\neg \mathcal{A}^{n}_{C}) = 0$.
	
	\noindent One theory that famously deals with the logical relations between categorical propositions is the \textit{theory of opposition} (henceforth: $\textit{TO}$). According to $\textit{TO}$, a set of logical relations holds between any propositions in every model: contrariety, contradiction, subcontrariety, and subalternation. Each of these logical relations can be defined in terms of constraints upon the truth-values of their relata. Thus, for any propositions $A$, $B$ and any model $w_{i}$: \newline
	
	\noindent $A$ and $B$ are contraries iff $v(w_{i}, A) = 1$ entails $v(w_{i}, B) = 0$.
	
	\noindent $A$ and $B$ are \textit{contradictories} iff $v(w_{i}, A) = 1$ entails $v(w_{i}, B) = 0$ and $v(w_{i}, A) = 0$ entails $v(w_{i}, B) = 1$.
	
	\noindent $A$ and $B$ are \textit{subcontraries} iff $v(w_{i}, A) = 0$ entails $v(w_{i}, B) = 1$.
	
	\noindent $B$ is the \textit{subaltern} of $A$ iff $v(w_{i}, A) = 1$ entails $v(w_{i}, B) = 1$ and $v(w_{i}, B) = 0$ entails $v(w_{i}, A) = 0$. \newline
	
	\noindent More generally, \cite{Hum13} rephrased all these logical relations into
	two main sets of a basic relation of entailment (symbol: $\vdash$): compatibility, and incompatibility. Opposition is initially a synonym for incompatibility,
	while subcontrariety and subalternation are further logical relations between compatible propositions. It concerns contrariety and contradiction and means that $A$ and $B$ are incompatible (symbol: $\bot$) iff these cannot be both true in the same model: \begin{center}
		
		$A\bot B =_{df} v(w_{i}, A) = 0$ iff $v(w_{i}, B) = 1.$
	\end{center}
	
	\noindent Or, equivalently, $A$ and $B$ are incompatible with each other iff the truth of $A$ in $w_{i}$ entails the falsity of $B$, i.e. the truth of $\neg B$:
	
	\begin{center}
		$A\bot B =_{df} A \vdash$$\neg B.$
	\end{center}
	
	\noindent Although $\textit{TO}$ was initially devoted to the logical relations between categorical
	propositions through Aristotle’s logical treatises, it has then be extended to any kind of propositional expressions. This theory consists in claiming that
	the aforementioned propositions stand in the same logical relations in any model whatsoever. Thus, \textit{TO} states that, for every $w_{i}$, the following holds in the set of Aristotelian categorical propositions, \textbf{X}: \newline
	
	\noindent $\textbf{A} \vdash \neg\textbf{E}, \textbf{A} \vdash \neg\textbf{0}, \neg\textbf{A} \vdash \textbf{O}, \textbf{A} \vdash \textbf{I}$,
	
	\noindent $\textbf{E} \vdash \neg\textbf{A}, \textbf{E} \vdash \neg\textbf{I}, \neg\textbf{E} \vdash \textbf{I}, \textbf{E} \vdash \textbf{O}$,
	
	\noindent $\textbf{I} \vdash \neg\textbf{E}, \neg\textbf{I} \vdash \neg\textbf{A},\neg\textbf{I} \vdash \textbf{E}, \neg\textbf{I} \vdash \textbf{O}$,
	
	\noindent $\textbf{O} \vdash \neg\textbf{A}, \neg\textbf{O} \vdash \textbf{I}$. \newline
	
	\noindent Exactly the same set of logical relations hold for their complementaries, i.e.
	the Keynesian categorical propositions, \textbf{X}$'$: \newline
	
	\noindent $\textbf{A}' \vdash \neg\textbf{E}', \textbf{A}' \vdash \neg\textbf{O}', \neg\textbf{A}' \vdash \textbf{O}', \textbf{A}' \vdash \textbf{I}'$,
	
	\noindent $\noindent \textbf{E}' \vdash \neg\textbf{A}', \textbf{E}' \vdash \neg\textbf{I}', \neg\textbf{E}' \vdash \textbf{I}', \textbf{E}' \vdash \textbf{O}'$,
	
	\noindent $\textbf{I}' \vdash \neg\textbf{E}', \neg\textbf{I}' \vdash \neg\textbf{A}',\neg\textbf{I}' \vdash \textbf{E}', \neg\textbf{I}' \vdash \textbf{O}'$,
	
	\noindent $\textbf{O}' \vdash \neg\textbf{A}', \neg\textbf{O}' \vdash \textbf{I}'$. \newline
	
	Now \cite{San68} recalled that \textit{TO} does not hold in every model but only in those models where the subject term A is not ‘empty’. An empty term being a predicate that applies to no individual, this means that \textit{TO} holds only in those models in which something instantiates $A$, i.e. every wi
	such that $v(w_{i}, Ax) = 1$. For let us suppose the contrary, i.e. a model $w_{j}$ in which nothing is $A$. Then $v(w_{j}, \exists x Ax) = 0$ and, hence, v$(w_{j}, \textbf{A}) = v(w_{j}
	, \textbf{E}) = 1$, and $v(w_{j}, \textbf{I}) = v(w_{j}, \textbf{O}) = 0$. This means that none of the logical relations of \textit{TO} holds in $wj$. And given that \textit{TO} is expected to hold in every model,
	\textit{TO} fails.
	
	\noindent A number of solutions have been proposed to revalidate \textit{TO}, and the related literature is famously referred to since medieval logic as the problem of existential import. According to their authors, some categorical propositions
	‘have’ import by naturally entailing the existence of their subject term:
	$(w_{i}, \mathcal{A}^{n}_{C} \vdash \exists xAx)$ for any $w_{i}$; whereas the other ones do not, or ‘lack’ import.
	Two main options are proposed in this respect: import by quality, (I$_{1}$), and import by quantity, (I$_{2}$). According to (I$_{1}$), the ‘affirmative’ propositions \textbf{A} and \textbf{I} always have import whereas \textbf{E} and \textbf{O} lack import. And according to (I$_{2}$), the ‘existential’ propositions \textbf{I} and \textbf{O} always have import (which accounts for how the traditional particular quantifier became the existential
	quantifier) whereas \textbf{A} and \textbf{E} lack import.
	
	\noindent Another problem raised by existential import is about logic as it stands: How
	can a theory be called ‘logical’ if it does not hold in \textit{every} model whatsoever?
	Note finally that the problem with (affirmative) existential import may also
	be extended in the form of a problem with ‘negative existential import’:
	although the literature about opposition has been entirely devoted to the
	logical relations between the set of Aristotelian propositions \{\textbf{A},\textbf{E},\textbf{I},\textbf{O}\}, \textit{TO}
	is also invalidated by the set of ‘complementary’ or Keynesian propositions \{\textbf{A}$'$,\textbf{E}$'$,\textbf{I}$'$,\textbf{O}$'$\}.
	For let us suppose now that there is a model $w_{k}$ in which \textit{everything} is $A$, so that the subject term $A$ may be said ‘full’ in $w_{k}$. Then the same
	logical trouble ensues in $w_{k}$, for $v(w_{k}, \textbf{A}') = v(w_{k}, \textbf{E}') = 1$ and $v(w_{k}, \textbf{I}') = v(w_{k}, \textbf{O}') = 0$. Hence \textit{TO} equally fails with affirmative and negative existential
	import, and the trouble goes beyond the sole case of ‘empty’ terms. It also concerns the predicate terms $B$, whose occurrence was even more neglected than the case in which everything is $A$; indeed, we will see later that the logical relations between Aristotelian and Keynesian categorical propositions also fails in models in which either nothing is $B$ or everything is $B$.
	
	Faced with all these difficulties, in the following we will propose a formal
	solution to the problem of existential import. That is: this problem does not
	stem from the informal reading of categorical propositions; rather, it relies on
	the way in which the latter are to be understood in light of their given logical
	forms. In other words, we want to restore the validity of \textit{TO} by showing that
	its alleged invalidity is due to an incorrect view about the logical form of $A^{4}_{C}$ and does not require any restriction on the models. For the corresponding
	theory would not be a logical theory properly speaking, otherwise.
	
	\subsection{Two Ways of Validity}
	
	There are two ways of constructing a semantics for \textit{TO}, in order to obtain
	a correspondence between a formal language and its model set. Either by
	restricting the available models in order to maintain the validity of some
	formulas, in accordance to their logical form. Or by altering these logical
	forms without imposing any restriction on their corresponding models. Let
	us call ‘model-adaptive strategy’ the first semantic method by virtue of which
	models are restricted in order to validate some logical forms. In contrast, let
	us call ‘language-adaptive strategy’ the second semantic method by virtue of
	which logical forms are adapted in order to be valid in unrestricted models.
	Although the latter view is the only one that seems to yield a properly ‘logical’
	theory and that will be endorsed in the rest of the paper accordingly, let us
	consider how to construct a model-adaptive semantics in the next section,
	in order to validate \textit{TO} in light of the historical tradition. The later section
	will be devoted to the language-adaptive method, where the solution relies
	on alternative logical forms for categorical propositions.

	\subsubsection{Semantics for restricted models (model-adaptive strategy)}
	
	The square of opposition is usually or ‘normally’ said to be valid in non-empty models, i.e. models in which what is predicated exists. Let us call ‘normal’ these restricted models that make \textit{TO} valid, accordingly. We make
	use of a \textit{bitstring semantics} for this purpose, in which all logical relations
	between arbitrary propositions is determined by a complete set of exclusive
	and exhaustive models in which they hold or not. These complementary
	models include submodels, and a model-adaptive semantics consists in constructing a set of such models by means of formulas whose logical relations are expected
	to hold. In the case of Aristotelian categorical propositions, it has been
	said previously (on page 7) that a number of logical dependencies are supposed to hold between the closed sets of Aristotelian and Keynesian categorical propositions. These logical relations appear in the following table where,
	for any formulas $A$, $B$: \newline
	
	\noindent $A \vdash B$ means that every model of $A$ is a model of $B$.
	
	\noindent $A \dashv B$ means that every model of $B$ is a model of $A$.
	
	\noindent $A \dashv\vdash B$ means that every model of $A$ is a model of $B$ and every model of $B$ is a model of $A$.
	
	\noindent $A \bot B$ means that every model of $A$ is not a model of $B$ and every model of $B$ is not a model of $A$.
	
	\begin{center}
		$\begin{array}
			{|c|c|c|c|c|} \hline & \textbf{A} & \textbf{E} & \textbf{I} & \textbf{O} \\ \hline
			\textbf{A} & \dashv\vdash & \bot & \vdash & \bot \\ \hline 
			\textbf{E} & \bot & \dashv\vdash & \bot & \vdash \\ \hline
			\textbf{I} & \dashv  & \bot & \dashv\vdash & \\ \hline
			\textbf{O} & \bot & \dashv & & \dashv\vdash \\ \hline
		\end{array}$  \end{center} 
	
	\noindent The above two empty boxes correspond to the relation of subcontrariety
	between \textbf{I} and \textbf{O}: there is a logical dependence between the latter, insofar as the falsehood of either one entails the truth of the other; however, this dependence is a case of compatibility in which not every model of a proposition
	puts any special constraint by excluding the other. For this reason, \textbf{I} and \textbf{O}
	may belong or not to one and the same model and their conjunction \textbf{I} $\wedge$ \textbf{O} holds there.
	
	\noindent The model-adaptive method \textit{TO} relies on the following tenets: the construction
	of a model consists in finding a set of formulas that are complementary with each other, i.e. incompatible with each other and not entailed by each other; a complete model for a given set of formulas is established once each such formula either entails or is incompatible with the other ones; any empty box of the table means that the corresponding formulas are compatible with each other and their conjunction holds. Each complete set of formulas is called a
	\textit{bit}, and the ordered combination of bits yields a \textit{bitstring} that characterizes a complete model for a given set of formulas.
	
	\noindent In the above table of logical relations, only \textbf{A} and \textbf{E} either entail or are
	incompatible with any other formula of the set of Aristotelian categorical propositions, \textbf{X}. Hence \textbf{A} and \textbf{E} correspond to two bits, i.e. two formulas for which distinct models are required. \textbf{I} and \textbf{O} are not bits, but their conjunction \textbf{I} $\wedge$ \textbf{O} might be so. Let us construct a further table to check the resulting
	logical relations.
	
	\begin{center}
		$\begin{array}
			{|c|c|c|c|} \hline & \textbf{A} & \textbf{E} & \textbf{I} \wedge \textbf{O} \\ \hline
			\textbf{A} & \dashv\vdash & \bot & \bot \\ \hline 
			\textbf{E} & \bot & \dashv\vdash & \bot \\ \hline
			\textbf{I} \wedge \textbf{O} & \bot & \bot & \dashv\vdash \\ \hline
		\end{array}$  
	\end{center} 
	
	\noindent All the above formulas are incompatible with each other, so the table is closed. It results in a set of 3 bits characterizing the meaning of any formula
	of $A$, i.e. the set of its truth-possibilities in terms of models. Thus, for any \textbf{X}
	in a ‘normal’ model set $W^{N}$ that validates $\mathbf{TO}$, its characteristic bitstring
	
	\begin{center}
		$v(W^{N},\textbf{X}) = \langle v(w_{1}^{N},\textbf{X}), v(w_{2}^{N},\textbf{X}), v(w_{3}^{N},\textbf{X})\rangle$
	\end{center}
	
	\noindent and corresponds to an ordered set of 3 complementary sets that exhaust the
	set of truth-possibilities in accordance to \textit{TO}: $w_{1}^{N}$ is the set in which \textbf{A} holds; $w_{2}^{N}$ is the set in which \textbf{I} $\wedge$ \textbf{O} holds; and $w_{3}^{N}$ is the set in which \textbf{E} holds. An arbitrary formula $\mathcal{A}$ holds in a model $w_{i}$ iff $v(w_{i}, A) = 1$, and it does not iff $v(w_{i}, A) = 0$. Here is a table that lists the meaning of Aristotelian categorical propositions in accordance to \textit{TO}. 
	
	\begin{center}
		$\begin{array}
			{|c|c|} \hline \textbf{X} & v(W^{N},\mathcal{A}^{4}_{C_{A}}) \\ \hline
			\textbf{A} & 100 \\ \hline 
			\textbf{E} & 001  \\ \hline
			\textbf{I} & 110  \\ \hline
			\textbf{O} & 011  \\ \hline
		\end{array}$  
	\end{center} 
	
	\noindent The ordered bits that symbolize the meaning of formulas helps to show their
	logical relations: every model of \textbf{A} is also a model of \textbf{I}, insofar as every
	ordered 1-bit of the former is also a 1-bit of the latter. This helps to rephrase
	logical relations between any formulas $A$, $B$ in those Boolean terms: \newline
	
	\noindent $A \bot B$ means that every ordered 1-bit of A is a 0-bit of B.
	
	\noindent $A \vdash B$ means that every ordered 1-bit for $A$ is a 1-bit for $B$.
	
	\noindent The same semantics obtains with the set of Keynesian categorical propositions
	\textbf{X}$'$, given that the set of logical relations between its corresponding formulas is identical to those between the Aristotelian categorical propositions. This
	leads to the isomorphic bitstring representation \begin{center}
		$v(W^{N},\textbf{X}') = \langle v(w_{1}^{N},\textbf{X}'), v(w_{2}^{N},\textbf{X}'), v(w_{3}^{N},\textbf{X}')\rangle$
	\end{center}
	
	\noindent where $w_{1}^{N}$ is the model set in which \textbf{A}' holds; $w_{2}^{N}$ is the model set in which \textbf{I}' $\wedge$ \textbf{O}' holds; and $w_{3}^{N}$ is the model set in which \textbf{E}$'$ holds. Whilst the same bitstrings obtain with \textbf{X}$'$, 
	
	\begin{center}
		$\begin{array}
			{|c|c|} \hline \textbf{X}' & v(W^{N},\mathcal{A}^{4}_{C_{A}}) \\ \hline
			\textbf{A}' & 100 \\ \hline 
			\textbf{E}' & 001  \\ \hline
			\textbf{I}' & 110  \\ \hline
			\textbf{O}' & 011  \\ \hline
		\end{array}$  
	\end{center} 
	
	\noindent the situation is different with the common set of \textit{interrelations} between Aristotelian
	and Keynesian categorical propositions, whose corresponding model includes
	their 4 + 4 = 8 formulas. 
	
	\begin{center}
		$\begin{array}
			{|c|c|c|c|c|c|c|c|c|} \hline & \textbf{A} & \textbf{E} & \textbf{I} & \textbf{O} & \textbf{A}' & \textbf{E}' & \textbf{I}' & \textbf{O}' \\ \hline
			\textbf{A} & \dashv\vdash & \bot & \vdash & \bot & & \bot & \vdash & \\ \hline 
			\textbf{E} & \bot & \dashv\vdash & \bot & \dashv & \bot & & & \vdash \\ \hline
			\textbf{I} & \dashv & \bot & \dashv\vdash & \bot & \dashv & & & \\ \hline
			\textbf{O} & \bot & \dashv & & \dashv\vdash & & \dashv & & \\ \hline
			\textbf{A}' & & \bot & \vdash & & \dashv\vdash & \bot & \vdash & \bot \\ \hline 
			\textbf{E}' & \bot & & & \vdash & \bot & \dashv\vdash & \bot & \vdash \\ \hline 
			\textbf{I}' & \dashv & & & & \dashv & \bot & \dashv\vdash & \\ \hline 
			\textbf{O}' & & \dashv & & & \bot & \dashv & & \dashv\vdash \\ \hline 
		\end{array}$  
	\end{center} 
	
	\noindent Again, empty boxes mean that their corresponding relata $A$, $B$ may occur in the same model consistently; these relata are replaced in a new whole table
	by their conjunction $A \wedge B$ in a new table of whole logical relations, until a
	final table that includes only incompatible sets of propositions. The process
	leads hereby to a final set of 7 incompatible ‘normal’ models for general
	categorical propositions $\mathcal{A}^{4}_{C} = \{\textbf{X}, \textbf{X}'\}$, instead of the previous 3 ones for
	strictly Aristotelian and strictly Keynesian categorical propositions:
	
	\begin{center}
		$v(W^{N},\mathcal{A}^{4}_{C}) = \langle v(w_{1}^{N},\textbf{X}'), ..., v(w_{7}^{N},\textbf{X}')\rangle$
	\end{center}
	
	\noindent where $w_{1}^{N}$ is the model set in which \textbf{A} $\wedge$ \textbf{A}$'$ holds; $w_{2}^{N}$ is the model set in
	which \textbf{A} $\wedge$ \textbf{O}$'$ holds; $w_{3}^{N}$ is the model set in which \textbf{A}$'$ $\wedge$ \textbf{O} holds; $w_{4}^{N}$ is
	the model set in which \textbf{I} $\wedge$ \textbf{O} $\wedge$ \textbf{I} $\wedge$ \textbf{O}$'$ holds; $w_{5}^{N}$ is the model set in which \textbf{I} $\wedge$ \textbf{E}$'$ holds; $w_{6}^{N}$ is the model set in which \textbf{E} $\wedge$ \textbf{I}$'$ holds; and $w_{7}^{N}$ is the model set in which \textbf{E} $\wedge$ \textbf{E}$'$ holds. 
	
	\noindent The set of ensuing logical relations between Aristotelian and Keynesian categorical
	propositions can be gathered from their following corresponding bitstrings:
	
	\begin{center}
		$\begin{array}
			{|c|c|} \hline \textbf{X} & v(W^{N},\mathcal{A}^{4}_{C}) \\ \hline
			\textbf{A} & 1100000 \\ \hline 
			\textbf{E} & 0000011  \\ \hline
			\textbf{I} & 1111100  \\ \hline
			\textbf{O} & 0011111  \\ \hline
			\textbf{A}' & 1010000 \\ \hline 
			\textbf{E}' & 0000101 \\ \hline 
			\textbf{I}' & 1111010 \\ \hline 
			\textbf{O}' & 0101111 \\ \hline 
		\end{array}$  
	\end{center} 
	
	\noindent where the relation of logical \textit{independence} is rendered in Boolean terms as
	follows: any two formulas $A, B$ are logically independent from each other iff not every model for $A$ is a model for either $B$ or $\neg A$, that is, i.e. not every
	ordered 1-bit for $A$ corresponds to either a 1-bit or a 0-bit for $B$.

	\subsubsection{Semantics for unrestricted models (language-adaptive strategy)}
	
	A second way to validate all the expected logical relations between categorical propositions is by altering their logical forms, instead of restricting models.
	Given that the central difficulty for \textit{TO} traditionally turns around existential
	import, the characteristic normal form of $\mathcal{A}^{4}_{C}$ should take this into account by introducing
	two additional clauses: that something is $A$; and that something is $B$. Nonetheless,
	it has been shown that other logical difficulties arise with what we called a ‘universal import’, i.e. whenever everything is either $A$ or $B$ in a model.
	Accordingly, the most comprehensive logical form representing all kinds of categorical propositions is \begin{center}
		$\pm\exists x$$\pm Ax \wedge \pm\exists x$$\pm Bx \wedge \pm\exists x(\pm Ax \wedge \pm Bx)$
	\end{center}
	
	\noindent The above extended logical form includes 7 literals, hence a resulting set
	of $2^{7} = 128$ categorical propositions. These correspond to a set of $8 \times 16$
	propositions, each of the initial 8 Aristotelian and Keynesian categorical
	propositions that are expressed by the third conjunct $\pm\exists x(\pm$$Ax \wedge \pm$$Bx)$ (and
	its 3 literals $\pm\exists x$, $\pm$$Ax$, $\pm$$Bx$) being now partitioned into $n = 16$ expressions
	of ontological commitment (about $A$, $\pm\exists x\pm$$Ax$, or about $B$, $\pm\exists x\pm$$Bx$) that
	are expressed by the first two conjuncts (and their 4 literals).
	
	Is this new logical form in position to revalidate all logical relations
	between Aristotelian categorical propositions, Keynesian categorical propositions,
	and the interrelations between both? The main trouble is with contradictory
	relations, because the latter only holds with an arbitrary proposition $A$ whose entire logical form is either affirmed, +$A$, or denied, $-A$. Thus affirming or denying each of the above literals is not enough to express contradictory
	relations between each of the available 128 categorical propositions.
	
	\noindent A way to revalidate the logical relations between Aristotelian categorical
	propositions has been proposed in \cite{Cha13}, however, and we
	will extend the proposed rationale from affirmative to negative existential
	import.
	
	\noindent According to this rationale, there are 3 kinds of Aristotelian categorical
	propositions with respect to the issue of existential import about $A$: \newline
	
	\noindent (1) those where having existential import is made explicit in the proposition:
	
	\noindent $\exists x Ax \wedge \mathcal{A}^{4}_{C}$ \newline
	
	\noindent (2) those where lacking existential import is made explicit in the proposition:
	
	\noindent $\neg\exists x Ax \wedge \mathcal{A}^{4}_{C}$ \newline
	
	\noindent (3) those where having existential import is left implicit:
	
	\noindent $\neg(\exists x Ax \wedge \neg \mathcal{A}^{4}_{C}$) \newline
	
	\noindent The latter case means the following: for any $A^{4}_{C}$, either existential import
	fails in it or nothing is said about it. This amounts to the disjunction $\neg\exists xAx
	\vee \mathcal{A}^{4}_{C}$, which is equivalent to $\neg(\exists x Ax \wedge \neg \mathcal{A}^{4}_{C})$. Given that (2) mostly leads to either inconsistencies or redundancies, the logical form of Aristotelian
	categorical propositions should be reduced to those relevant cases (1)-(3),
	thereby turning the first literal $\pm\exists x$ into a fixed component: \begin{center}
		
		$\mathcal{A}^{4}_{C_{\pm A\pm B}} = \exists x\pm Ax \wedge \pm\exists x\pm Bx \wedge \pm\exists x(\pm Ax \wedge \pm Bx)$ \end{center}
	
	\noindent There are only $2^{6} = 64$ relevant categorical propositions with the remaining 6 literals, accordingly. In order to streamline the symbolization of categorical propositions for
	the rest of the paper, let us rewrite these as follows. For every predicate term $P = \{\pm$$A, \pm$$B\}$: \newline
	
	\noindent $\mathcal{A}^{4}_{C_{P!}}$ is the set of categorical propositions affirming the existence of $\pm$$Ps$:
	
	$\mathcal{A}^{4}_{C_{P!}} = \exists x\pm$$Px \wedge A^{4}_{C}$
	
	\noindent $\mathcal{A}^{4}_{C_{P?}}$
	is the set of categorical propositions not affirming the existence of $\pm$$Ps$:
	
	$\mathcal{A}^{4}_{C_{P?}} = \neg(\exists x\pm$$Px \wedge \neg A^{4}_{C})$ \newline
	
	\noindent The same process can be applied with respect to the issue of ‘negative existential import’ (whenever something is not-$A$), i.e. the set of categorical
	propositions $\mathcal{A}^{4}_{C_{\overline{A!}}}$. This yields a corresponding table of formulas for Keynesian
	categorical propositions, where the symbols from \cite{Cha13} and
	\cite{Rea15} are reworded correspondingly.
	
	\noindent To summarize: formulas with explicit affirmative existential import are cases
	in which categorical propositions have import about A and are thereby
	limited to $6 - 1 = 5$ literals: 
	\begin{center}
		$\mathcal{A}^{4}_{C_{A!}} = \exists x Ax \wedge \pm\exists x\pm$$Bx \wedge \pm\exists x(\pm$$Ax \wedge \pm$$Bx)$
	\end{center}
	
	\noindent By extension, propositions with explicit negative existential import about $A$
	are those which have import about not-$A$:
	\begin{center}
		$\mathcal{A}^{4}_{C_{\overline{A!}}} = \exists x\neg Ax \wedge \pm\exists x\pm$$Bx \wedge \pm\exists x(\pm$$Ax \wedge \pm$$Bx)$
	\end{center}
	
	\noindent A generalized theory of categorical propositions consists in applying the same
	process to the 16 cases of ontological commitment about $A$ and $B$, where the
	traditional issue of (affirmative) existential import includes 4 kinds of these
	logical forms.
	
	\noindent \textit{TO} is restored once a distinction is made between categorical propositions
	with or without explicit import, whereas the failure of \textit{TO} arises where any of
	the four related propositions lacks explicit import. The result is that, instead of being invalidated, \textit{TO} includes 3 valid logical squares with extended logical
	forms (with explicit or implicit import): \newline
	
	\noindent $S_{1}(A^{4}_{C}) = \{\textbf{A}_{A!}, \textbf{E}_{A!}, \textbf{I}_{A?}, \textbf{O}_{A?}\}$
	
	\noindent $S_{2}(A^{4}_{C}) = \{\textbf{A}_{A!}, \textbf{E}_{A?}, \textbf{I}_{A!}, \textbf{O}_{A?}\}$
	
	\noindent $S_{3}(A^{4}_{C}) = \{\textbf{A}_{A?}, \textbf{E}_{A!}, \textbf{I}_{A?}, \textbf{O}_{A!}\}$ \newline
	
	\noindent 3 complementary squares follow from such formulas and their characteristic
	bitstrings, isomorphically to the ones for Aristotelian categorical propositions:
	
	\noindent $S_{1}(\mathcal{A}^{4}_{C}) = \{\textbf{A}_{\overline{A!}}, \textbf{E}_{\overline{A!}}, \textbf{I}_{\overline{A?}}, \textbf{O}_{\overline{A?}}\}$
	
	\noindent $S_{2}(\mathcal{A}^{4}_{C}) = \{\textbf{A}_{\overline{A!}}, \textbf{E}_{\overline{A?}}, \textbf{I}_{\overline{A!}}, \textbf{O}_{\overline{A?}}\}$
	
	\noindent $S_{3}(\mathcal{A}^{4}_{C}) = \{\textbf{A}_{\overline{A?}}, \textbf{E}_{\overline{A!}}, \textbf{I}_{\overline{A?}}, \textbf{O}_{\overline{A!}}\}$ \newline
	
	\noindent The logical relations holding in these 3 squares may be checked by finding the
	characteristic bitstrings of their relata. An unrestricted model for any categorical proposition $\mathcal{A}^{4}_{C}$ is an ordered set of $2^{4} = 16$ model sets relating $n = 2$ classes $A$ and $B$, such that there can be a combination of 4 possible relations $(i)$--$(iv)$ between elements of $A$ and $B$: $(i)$ some $A$ is $B$, $(ii)$ some $A$ is not-$B$; $(iii)$ some not-$A$ is $B$; some not-$A$ is not-$B$. The resulting 16 model sets are the following: \newline
	
	$w_{1} = \{(i),(ii),(iii),(iv)\}$ \hspace{5cm} $w_{9} = \{(i),(iv)\}$
	
	$w_{2} = \{(i),(ii),(iii)\}$
	\hspace{5,72cm} $w_{10} = \{(ii)(,iv)\}$
	
	$w_{3} = \{(i),(ii),(iv)\}$
	\hspace{5,78cm} $w_{11} = \{(iii),(iv)\}$
	
	$w_{4} = \{(i),(iii),(iv)\}$
	\hspace{5,68cm} $w_{12} = \{(i)\}$
	
	$w_{5} = \{(ii),(iii),(iv)\}$
	\hspace{5,58cm} $w_{13} = \{(ii)\}$
	
	$w_{6} = \{(i),(ii)\}$
	\hspace{6,52cm} $w_{14} = \{(iii)\}$
	
	$w_{7} = \{(i),(iii)\}$
	\hspace{6,42cm} $w_{15} = \{(iv)\}$
	
	$w_{8} = \{(ii),(iii)\}$
	\hspace{6,32cm} $w_{16} = \{ \}$ \newline
	
	\noindent It follows from it that the ‘internal’ logical relations between
	Aristotelian categorical propositions hold after adding the clause of affirmative
	existential import about $A$ in their logical form. The same does for the
	‘internal’ relations between Keynesian categorical propositions, after assuming
	negative existential import about $A$. At the same time, the logical interrelations
	between Aristotelian and Keynesian categorical proposition still do not hold in some models. No wonder, since these interrelations hold only if existential assumption
	is made both about $A$ and $B$. Whilst \cite{Cha13} and \cite{Rea15} dealt only with Aristotelian categorical propositions and existential
	import about $A$, a complete survey of the logical forms validating these
	interrelations requires a combination of 4 possible existential assumptions:
	affirmative and negative, on the one hand; about $A$ or $B$, on the other hand.

	For a comparative analysis of the categorical propositions that validate or
	invalidate \textit{TO}, a comprehensive list of the 16 kinds of existential commitment
	can be made about their two predicate terms. The resulting 256 propositions,
	i.e. the $16 \times 8 = 128$ categorical propositions with explicit import $\mathcal{A}^{4}_{P!}$
	together with their 128 contradictories $\mathcal{A}^{4}_{P?}$, may include from 0 to 4 kinds
	of existential assumption, whilst the aforementioned ‘universal’ assumptions (i.e. that either everything or nothing is $P$) merely amount to a lack of
	existential assumption. The resulting number of valid square is considerably
	extended, assuming that any proper square must include a set of four categorical
	propositions with two pairs of universals-existentials and affirmatives-negatives
	fulfilling logical requirements.

	\section{Conclusion and Prospects}
	
	Let us recapitulate the content of the present paper. Our aim was to extend the usual definition of categorical propositions in order to emphasize their
	logical form in terms of literals (both in TM and FOL).
	A generalized solution to the problem of existential import has been proposed
	beyond one former proposal by \cite{Cha13}, including an introduction
	into a universal import for Keynesian propositions.
	The resulting semantics of bitstrings can be extended to any kinds of formulas, beyond the present of categorical propositions. An interesting case study is
	the set of knowledge statements considered in \cite{Eng69}, whose general logical form
	
	\begin{center}
		$\pm@\pm$$K\pm$$p$
	\end{center}
	
	\noindent may be treated in the same pattern without introducing any intensional account of modal semantics. Another case is the set of dyadic relations, whose
	binary predicates of the form 
	
	\begin{center}
		$\pm\exists x$$\pm\exists y\pm$$Rxy$
	\end{center}
	
	\noindent can also be analyzed in a purely Boolean way.
	
	\noindent Finally, the present paper needs some automatic process in order to determine
	the nature of logical relations between any pair of the available 256 categorical
	propositions. This requires the implementation of a programming machine in
	the style of Prolog, and this project will be pursued along with J.-Mart\'{i}n
	Castro-Manzano.
	\nocite{*}
	\bibliographystyle{eptcs}
	\bibliography{biblio}
	
\end{document}